\documentclass{article}

\usepackage{amssymb}
\usepackage{graphicx}
\usepackage{multirow}
\usepackage{url}
\usepackage{listings}
\usepackage{color}
\usepackage{fRReE}
\usepackage[utf8]{inputenc}
\usepackage[T1]{fontenc}
\usepackage{lmodern}
\usepackage{url}
\usepackage[english]{babel}
\usepackage{amsmath}
\usepackage{amssymb,amsfonts,textcomp}
\usepackage{color}
\usepackage{array}
\usepackage{hyperref}

\RRlogo{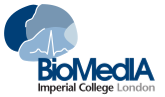}

\RRdate{July 2015}

\RRauthor{
Romain \textsc{Reuillon}
\RRaffiliation[iscpif]{Institut des Syst\`emes Complexes 
Paris, \^Ile de France}
\and
\\Mathieu \textsc{Leclaire}\RRaffiliationref{iscpif}
\and
\\Jonathan \textsc{Passerat-Palmbach}
\RRaffiliation[imperial]{Imperial College London, Department of Computing, London SW7 2AZ}
}

\RRauthorhead{Reuillon et. al}
\RRtitlehead{HPCS 2015}

\RRsource{IEEE High Performance Computing and Simulation conference 2015}
\RRcopyright{\textcopyright 2015 IEEE}
\RRoriginalpages{n/a-n/a}
\RRadditionalinformation{to be published}

\RRtitle{\huge{Model Exploration Using OpenMOLE}\\ a workflow engine for large scale distributed design of experiments and parameter tuning}

\RRabstract{
  OpenMOLE is a scientific workflow engine with a strong emphasis on workload distribution.
Workflows are designed using a high level Domain Specific Language (DSL) built on top of Scala. It exposes natural parallelism constructs to easily delegate the workload resulting from a workflow to a wide range of distributed computing environments. In this work, we briefly expose the strong assets of OpenMOLE and demonstrate its efficiency at exploring the parameter set of an agent simulation model. We perform a multi-objective optimisation on this model using computationally expensive Genetic Algorithms (GA). OpenMOLE hides the complexity of designing such an experiment thanks to its DSL, and transparently distributes the optimisation process. The example shows how an initialisation of the GA with a population of 200,000 individuals can be evaluated in one hour on the European Grid Infrastructure.

}

\RRkeyword{Distributed computing; Scientific workflows; Model Exploration; Design of Experiments; Evolutionary Algorithms}

\begin{document}

\lstset{
        frame=none,
        linewidth=0.97\columnwidth,
        xleftmargin=6pt,
        columns=flexible,
        keywordstyle=\color{blue}\bfseries, 
        identifierstyle=\ttfamily, 
        commentstyle=\color{blue}, 
        stringstyle=\ttfamily\color{red}, 
        numbers=none,
        stepnumber=1,
        numberfirstline=true,
        numbersep=4pt,
        rangeprefix=//*,
        rangesuffix=*//,
        includerangemarker=false,
        breaklines = true,
        captionpos=b,
}

\makeRR

\lstdefinelanguage{scala}{
  morekeywords={abstract,case,catch,class,def,
    do,else,extends,false,final,finally,
    for,if,implicit,import,match,mixin,
    new,null,object,override,package,
    private,protected,requires,return,sealed,
    super,this,throw,trait,true,try,
  type,val,var,while,with,yield},
  otherkeywords={=>,<-,<\%,<:,>:,\#,@},
    sensitive=true,
    morecomment=[l]{//},
    morecomment=[n]{/*}{*/},
    morestring=[b]",
    morestring=[b]',
    morestring=[b]"""
  }

\section{\textbf{Introduction}}
{\color{black}
Parameter tuning is a daily problem in any scientific community using
complex algorithms. In the specific case of simulation applications,
increased p\foreignlanguage{english}{erformance of computer
architectures have led to more and more ambitious
}\foreignlanguage{english}{models}\foreignlanguage{english}{
}\foreignlanguage{english}{with a growing number of parameters.
}\foreignlanguage{english}{Therefore
}\foreignlanguage{english}{exploring high dimensional spaces
to}\foreignlanguage{english}{
tun}\foreignlanguage{english}{e}\foreignlanguage{english}{ parameters
for specific problems }\foreignlanguage{english}{has
}\foreignlanguage{english}{become a
}\foreignlanguage{english}{central}\foreignlanguage{english}{ problem.
}\foreignlanguage{english}{Stochastic
}\foreignlanguage{english}{models}\foreignlanguage{english}{ add
another dimension of parameters to explore,
a}\foreignlanguage{english}{s}\foreignlanguage{english}{
}\foreignlanguage{english}{different random sources
}\foreignlanguage{english}{should generally
}\foreignlanguage{english}{be
}\foreignlanguage{english}{tested}\foreignlanguage{english}{ for each
set of parameters, in order to obtain statistically sound
}\foreignlanguage{english}{results}\foreignlanguage{english}{.}}

{\color{black}
This wide range of parameters to tune, combined with the intrinsic
execution time of the application, make it impossible to run significant
Designs of Experiments (DoE) on a desktop computer. Experiments within
a DoE can be processed independently from each other. They are perfect
candidates for distributed computing platforms. In an ideal world, the
time required to process the whole DoE would be almost equivalent to
the execution time of a single experiment. However the methodological
and technical costs of using distributed execution environments imply
that most parameter space explorations are achieved either on a single
desktop computer and occasionally on a multi-core server with shared
memory. Larger scale platforms are rarely used, although clusters or
worldwide computing infrastructures like EGI (European Grid Initiative)
are well suited for this kind of applications.}

{\color{black}
Compared to other workflow processing engines, OpenMOLE promotes a
zero-deployment approach by accessing the computing environments from
bare metal, and copies on-the-fly any software component required for a
reliable remote execution. OpenMOLE also encourages the use of software
components developed in heterogeneous programming languages and enables
users to easily replace the elements involved in the workflow.
Workflows can be designed using either a Graphical User Interface (GUI), or a Domain Specific
Language (DSL) which exposes advanced workflow design constructs.}

{\color{black}
Apart from these core elements of the platform, OpenMOLE ships with its
own software ecosystem. It contains among others
GridScale\footnote{\url{https://github.com/openmole/gridscale}}, a
library to access a wide range of computing environment, and
Yapa\footnote{\url{https://github.com/openmole/yapa}}, a packaging tool
ensuring the successful re-execution of applications across
heterogeneous platforms. For more details regarding the core
implementation and features of OpenMOLE, interested readers can
refer to \cite{Reuillon.etal.2010, Reuillon.etal.2013} and the
OpenMOLE website\cite{Reuillon.etal.2015}.

{\color{black}
OpenMOLE focuses on making distributed computing available in the most
straightforward way to the scientific community. Depending on the
applications, several problems might arise when attempting to
distribute the execution. Software tools can help scientists overcome
these barriers. This paper describes the input of OpenMOLE and its
software ecosystem to the distribution of complex scientific
applications to remote execution environments.}

{\color{black}
We first detail the problems faced by the scientific community to
effectively distribute an application. Then, we present how the tools
from the OpenMOLE ecosystem can answer these problems. The last section
presents a test case showing how an actual simulation application was
successfully distributed using OpenMOLE.}

\section{\textbf{What is OpenMOLE?}}
{\color{black}
OpenMOLE is a scientific workflow engine with facilities to delegate
its workload to a wide range of distributed computing environments. It shows
several main advantages with respect to the other workflow management
tools available. First, OpenMOLE distinguishes as a tool that does not
target a specific scientific community, but offers generic tools to
explore large parameter sets.}

{\color{black}
Second, OpenMOLE features a Domain Specific Language (DSL) to
describe the workflows. According to \cite{Barker2008}, workflow platforms
should not introduce new languages but rely on established ones.
OpenMOLE{\textquotesingle}s DSL is based on the high level Scala
programming language \cite{Odersky2004}. In addition to the DSL, a web
interface is currently under development, and will permit expressing workflows
graphically.}

{\color{black}
Finally, OpenMOLE features a great range of platforms to distribute the
execution of workflows, thanks to the underlying GridScale
library\footnote{\url{https://github.com/openmole/gridscale}}.
GridScale is part of the OpenMOLE ecosystem and acts as one of its
foundation layers. It is responsible for accessing the different
execution environments. The last release of OpenMOLE can target SSH
servers, multiple cluster managers and computing grids ruled by the
gLite/EMI middleware.}

{\ttfamily\color{black}
\textrm{In this section, we describe t}\textrm{wo main components of the
}\textrm{OpenMOLE }\textrm{platform: the Domain Specific Language and
the distributed environments.}\textrm{ The}\textrm{y contribute to make
the}\textrm{ explor}\textrm{ation} \textrm{of a parameter set
}\textrm{simple to}\textrm{ distribute.}}

\subsection{\textbf{A DSL to describe workflows}}
{\color{black}
Scientific experiments are characterised by their ability to be
reproduced. This implies capturing all the processing stages leading to
the result. Many execution platforms introduce the notion of workflow
to do so \cite{Barker2008,Mikut2013}. Likewise, OpenMOLE manipulates
workflows and distribute their execution across various computing
environments.}

A workflow is a set of~tasks~linked with each other
through transitions. From a high level point of view,
tasks comprise~inputs,~outputs~and~optional default values. Tasks describe what OpenMOLE
should execute and delegate to remote environments. They embed the
actual applications to study. Depending on the kind of program (binary
executable, Java...) to embed in OpenMOLE, the user chooses the
corresponding task. Tasks execution~depends on inputs variables, which
are provided by the dataflow. Each task~produces outputs~returned
 to the dataflow and transmitted to the input of consecutive
tasks. OpenMOLE exposes several facilities to inject data in the
dataflow (\textit{sources}) and extract useful results at the end of the experiment (\textit{hooks}).

{\color{black}
Two choices are available when it comes to describe a workflow in
OpenMOLE: the Graphical User Interface (GUI) and the Domain Specific
Language (DSL). Both strategies result in identical workflows. They can
be shared by users as a way to reproduce their execution.}

{\color{black}
OpenMOLE{\textquotesingle}s DSL is based upon the Scala programming
language, and embeds new operators to manage the construction and
execution of the workflow. The advantage of this approach lies in the
fact that workflows can exist even outside the OpenMOLE environment. As
a high-level language, the DSL can be assimilated to an algorithm
described in pseudocode, understandable by most scientists. Moreover,
it denotes all the types and data used within the workflow, as well as
their origin. This reinforces the capacity to reproduce workflow
execution both within the OpenMOLE platform or using another tool.}

The philosophy of OpenMOLE is~\textit{test small}~(on your computer) and
\textit{scale for free}~(on remote distributed computing environments).
The DSL supports all the Scala constructs and provides additional operators and
classes especially designed to compose workflows. OpenMOLE workflows
expose explicit parallel aspects of the workload that can be delegated
to distributed computing environments in a transparent manner. The next
sections introduces the available computing environments.

\subsection{\textbf{Distributed Computing environments}}
{\color{black}
OpenMOLE helps delegate the workload to a wide range of HPC
environments including remote servers (through SSH), clusters
(supporting the job schedulers PBS, SGE, Slurm, OAR and Condor) and
computing grids running the gLite/EMI middleware.}

{\color{black}
Submitting jobs to distributed computing environments can be complex for
some users. This difficulty is hidden by the GridScale library from the
OpenMOLE ecosystem. GridScale provides a high level abstraction to all
the execution platforms mentioned previously.}

{\color{black}
When GridScale was originally conceived, a choice was made not to rely
on a standard API (Application Programming Interface) to interface with
the computing environments, but to take advantage of the command line
tools available instead. As a result, GridScale can embed any job
submission environment available from a command line. From a higher
perspective, this allows OpenMOLE to work seamlessly with any computing
environment the user can access.}

{\color{black}
Users are only expected to select the execution environment for the
tasks of the workflow. This choice can be guided by two considerations:
the availability of the resources and their suitability to process a
particular problem. The characteristics of each available environment
must be considered and matched with the application{\textquotesingle}s
characteristics. Depending on the size of the input and output data,
the execution time of a single instance and the number of independent
executions to process, some environments might show more appropriate
than others. }

{\color{black}
At this stage, OpenMOLE{\textquotesingle}s simple workflow description is quite convenient to determine the computing environment best
suited for a workflow. Switching from one environment to another is
achieved either by a single click (if the workflow was designed with
the GUI) or by modifying a single line (for workflows described using
the DSL).}

{\color{black}
Some applications might show more complicated than others to distribute.
The next section exposes the main challenges a user is faced with when
trying to distribute an application. We present how OpenMOLE couples
with a third-party software called CARE to solve these problems.}

\section{\textbf{The Challenges of Distributing Applications}}
\subsection{\textbf{Problems and classical solutions}}
{\color{black}
Let us consider all the dependencies introduced by software bundles
explicitly used by the developer. They can take various forms depending
on the underlying technology. Compiled binary applications will rely on
shared libraries, while interpreted languages such as Python will call
other scripts stored in packages.}

{\color{black}
These software dependencies become a problem when distributing an
application. It is indeed very unlikely that a large number of remote
hosts are deployed in the same configuration as a researcher’s
desktop computer. Actually, the larger the pool of distributed
machines, the more heterogeneous they are likely to be.}

{\color{black}
If a dependency is missing at runtime, the remote execution will simply
fail on the remote hosts where the requested dependencies are not
installed. An application can also be prevented from running properly
due to incompatibilities between versions of the deployed dependencies.
This case can lead to silent errors, where a software dependency would
be present in a different configuration and would generate different
results for the studied application.}

{\color{black}
Silent errors break Provenance, a major concern of the scientific
community \cite{Miles.etal.2007, MacKenzie-Graham.etal.2008}.
Provenance criteria are satisfied when an application is documented
thoroughly enough to be reproducible. This can only happen in
distributed computing environments if the software dependencies are
clearly described and available.}

{\color{black}
Some programming environments provide a solution to these problems.
Compiled languages such as C and C++ offer to build a static binary,
which packages all the software dependencies. Some applications can be
very difficult to compile statically. A typical case is an application
using a closed source library, for which only a shared library is
available.}

{\color{black}
Another approach is to rely on an archiving format specific to a
programming language. The most evident example falling into this
category are Java Archives (JAR) that embed all the Java libraries an
application will need.}

{\color{black}
A new trend coming from recent advances in the software engineering
community is embodied by Docker. Docker has become popular with DevOps
techniques to improve software developers efficiency. It enables them
to ship their application within a so-called container that will
include the application and its required set of dependencies.
Containers can be transferred just like an archive and re-executed on
another Docker engine. Docker containers run in a sandboxed virtual
environment but they are not to be confound with virtual machines. They
are more lightweight as they don{\textquotesingle}t embed a full
operating system stack. The use of Docker for reproducible research has
been tackled in \cite{Chamberlain.Schommer2014}.}

{\color{black}
The main drawback of Docker is that it implies deploying a Docker engine
on the target host. Having a Docker engine running on every target host
is a dodgy assumption in heterogeneous distributed environments such as
computing grids.}

{\color{black}
The last option is to rely on a third-party application to generate
re-executable applications. The strategy consists in collecting
all the dependencies during a first execution in order to store them in
an archive. This newly generated bundle is then shipped to remote hosts
instead of the original application. This is the approach championed by
tools like CDE \cite{Guo2012} or CARE \cite{Janin.etal.2014}.}

Considering all these aspects, the OpenMOLE platform has for long
chosen to couple with tools providing standalone packages. While CDE
was the initial choice, recent requirements in the OpenMOLE user
community have led the development team to switch to the more flexible
CARE. The next section will detail how OpenMOLE relies on CARE to
package applications.}

\subsection{\textbf{Combining OpenMOLE with CARE}}

The first step towards spreading the workload across heterogeneous
computing elements is to make the studied application executable on the
greatest number of environments. We have seen previously that this
could be difficult with the entanglement of complex software environments
available nowadays. For instance, a Python script will run only in a
particular version of the interpreter and may also make use of binary
dependencies. The best solution to make sure the execution will run as
seamlessly on a remote host as it does on the desktop machine of the
scientist is to track all the dependencies of the application and ship
them with it on the execution site.

OpenMOLE used to provide this feature through a third-party tool called CDE (Code,
Data, and Environment packaging) \cite{Guo2012}. CDE creates archives
containing all the items required by an application to run on any
recent Linux platform. CDE tracks all the files that interact with the
application and creates the base archive.

The only constraint regarding CDE is to create the archive on a
platform running a Linux kernel from the same generation as those 
of the targeted computing elements. As a rule of thumb, a good
way to ensure that the deployment will be successful is to create the 
CDE package from a system running
Linux 2.6.32. Many HPC environments run this version, as it is the
default kernel used by science-oriented Linux distribution, such as
Scientific Linux and CentOS.

CARE on the other hand presents more advanced features than CDE. CDE
actually displays the same limit than a traditional binary run on a remote
host: i.e. the archive has to be generated on a platform running an old
enough Linux kernel, to have a maximum compatibility with remote hosts.
CARE lifts this constraint by emulating missing system calls on the
remote environment. Thus, an application packaged on a recent release
of the Linux kernel will successfully re-execute on an older kernel
thanks to this emulation feature. Last but not least,
CDE{\textquotesingle}s development has been stalled over the last few
years, whereas CARE was still actively developed over the last few
months. CARE{\textquotesingle}s developers are also very reactive
when an eventual problem is detected in their piece of software. This
makes CARE a reliable long-term choice for re-execution facilities.
All that remains is to complete the package by adding specific
customisations related to the integration of the application within an
OpenMOLE workflow.

As previously evoked, OpenMOLE workflows are mainly composed of tasks.
Different types of tasks exist, each embedding a different kind of
application. Generic applications such as those packaged with CARE are
handled by the \textit{SystemExecTask}. As an OpenMOLE task, the
generated element is ready to be added to the OpenMOLE scene and
integrated in a workflow.

We will now demonstrate the use of the DSL and computing environments with a 
concrete example. For the sake of simplicity, we will
exploit a simulation model developed with the NetLogo\cite{Wilensky1999} platform. This 
simulation platform benefits from a native integration in OpenMOLE, which 
ensures the model will run on remote hosts. It spares the user from the extra
packaging step using CARE that was introduced in this section.

\section{\textbf{An A to Z example: Calibrating a model using Genetic
Algorithms}}
{\color{black}
This example presents step by step how to explore a NetLogo model with
an Evolutionary/Genetic Algorithm (EA/GA) in OpenMOLE.
We{\textquotesingle}ve chosen NetLogo for its simplicity to design
simple simulation models with a graphical output quickly. However, this
approach can be applied to any other kind of simulation model,
regardless of their implementation platform.}

\subsection{\textbf{The ant model}}

\textcolor{black}{We demonstrate this
example using the ants foraging
model present in the Netlogo library. This model was created by Ury
Wilensky. According
to~}\href{http://ccl.northwestern.edu/netlogo/models/Ants}{\textcolor{black}{NetLogo{\textquotesingle}s
website}}\textcolor{black}{, this model is described as:~}\newline
\textcolor{black}{“}\textit{\textcolor{black}{In this project, a colony
of ants forages for food. Though each ant follows a set of simple
rules, the colony as a whole acts in a sophisticated way. When an ant
finds a piece of food, it carries the food back to the nest, dropping a
chemical as it moves. When other ants “sniff” the chemical, they follow
the chemical toward the food. As more ants carry food to the nest, they
reinforce the chemical trail.”}}

\begin{figure}
\centering
\includegraphics[width=7.541cm,height=6.795cm]{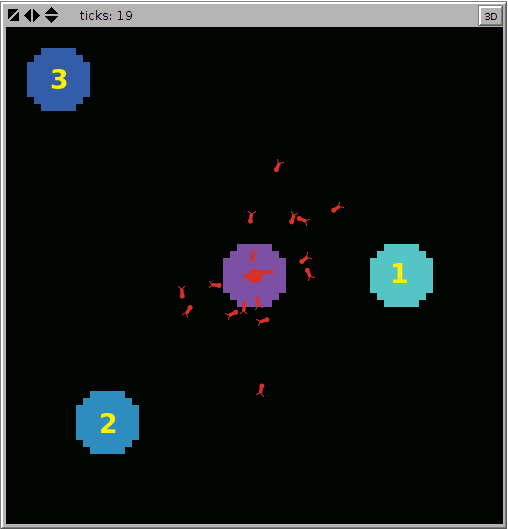}
\caption{\textbf{Visual Representation of the Ant Model Showing the 3 Food Sources and the Multiple Ant Agents}}
\label{fig:ant-model}
\end{figure}

A visual representation of this model appears in Figure \ref{fig:ant-model}.

\textcolor{black}{In this tutorial we use~}\textcolor{black}{a headless
version}\textcolor{black}{~}\textcolor{black}{of the model. This
modified version is availabl}\textcolor{black}{e from the
OpenMOLE{\textquotesingle}s website}\footnote{\url{http://www.openmole.org/current/ants.nlogo}}.

\vfill
\eject

\subsection{\textbf{Define the problem to solve as an optimisation problem}}

{\color{black}
This model manipulates three parameters:}

\begin{figure}
\centering
\includegraphics[width=6.227cm,height=6.466cm]{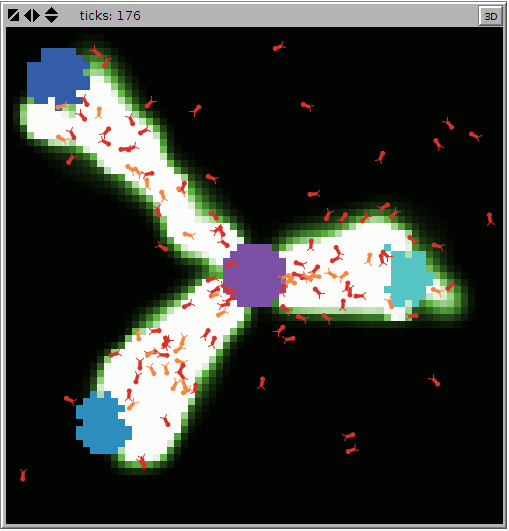}
\caption{\textbf{Graphical Output of the Ants Agents Foraging from 3 Different Sources}}
\label{fig:ants-foraging}
\end{figure}

\begin{itemize}
\item \textcolor{black}{\ }\textit{\textcolor{black}{Population}}\textcolor{black}{:
number of ants in the model,}
\item {\color{black}
\ \textit{Evaporation-rate}: controls the evaporation rate of the
chemical,}
\item {\color{black}
\textit{\ Diffusion-rate}: controls the diffusion rate of the chemical.}
\end{itemize}
{\color{black}
Ants forage from three sources of food as represented in Figure \ref{fig:ants-foraging}).
Each source is positioned at different distances from the ant colony.}

{\centering \par}

{\color{black}
In this example, we want to search the best combination of the two
parameters \textit{evaporation-rate} and \textit{diffusion-rate} which minimises the
eating time of each food source. We will use
OpenMOLE{\textquotesingle}s embedded Evolutionary Algorithms features
to perform this optimisation process. The first thing is to define a
fitness function describing the optimisation problem.}

{\color{black}
We build our fitness function by modifying the NetLogo Ants source code
to store for each food source the first ticks indicating that this food
source is empty, as shown in Listing \ref{lis:fitness}:}

\begin{lstlisting}[language=logo, label=lis:fitness, caption={\textbf{NetLogo Function Returning the Simulation Tick at Which each Food Source Became Empty}}]

to compute-fitness
  if ((sum [food] of patches with [food-source-number = 1] = 0) and (final-ticks-food1 = 0)) [
    set final-ticks-food1 ticks ]
  if ((sum [food] of patches with [food-source-number = 2] = 0) and (final-ticks-food2 = 0)) [
    set final-ticks-food2 ticks ]
  if ((sum [food] of patches with [food-source-number = 3] = 0) and (final-ticks-food3 = 0)) [
    set final-ticks-food3 ticks ]
end

\end{lstlisting}

{\color{black}
At the end of each simulation we return the values for the three
objectives:}

\begin{itemize}
\item {\color{black}
The simulation ticks indicating that source 1 is empty,}
\item {\color{black}
The simulation ticks indicating that source 2 is empty,}
\item {\color{black}
The simulation ticks indicating that source 3 is empty.}
\end{itemize}
\textcolor{black}{The combination of the three objectives indicates the
quality of the parameters used to run the simulation. This situation is
a~}\href{http://en.wikipedia.org/wiki/Multiobjective$_{}$optimization}{\textcolor{black}{multi-objective
optimisation}}\textcolor{black}{~}\textcolor{black}{problem. In case
there is a compromise between these three objectives, we will obtain
a~}\href{http://en.wikipedia.org/wiki/Pareto$_{}$efficiency}{\textcolor{black}{Pareto
frontier}}\textcolor{black}{~}\textcolor{black}{at the end of the
optimisation process.}

\subsection{\textbf{Getting the ant model to run in OpenMOLE}}
{\color{black}
When building a calibration or optimisation workflow, the first step is
to make the model run in OpenMOLE. The script displayed in Listing \ref{lis:openmole-workflow} simply embeds the
NetLogo model and runs one single execution of the model with arbitrary
parameters.}

\begin{lstlisting}[language=scala,label=lis:openmole-workflow, caption={\textbf{Complete OpenMOLE Workflow Embedding the Ant Model}}]

// Define the input variables
val gPopulation = Val[Double]
val gDiffusionRate = Val[Double]
val gEvaporationRate = Val[Double]
val seed = Val[Int]

// Define the output variables
val food1 = Val[Double]
val food2 = Val[Double]
val food3 = Val[Double]

// Define the NetlogoTask
val cmds = Seq("random-seed ${seed}", "run-to-grid")
val ants =
  NetLogo5Task("Ants.nlogo", cmds) set (
    // Map the OpenMOLE variables to NetLogo variables
    netLogoInputs += (gPopulation, "gpopulation"),
    netLogoInputs += (gDiffusionRate, "gdiffusion-rate"),
    netLogoInputs += (gEvaporationRate, "gevaporation-rate"),
    netLogoOutputs += ("final-ticks-food1", food1),
    netLogoOutputs += ("final-ticks-food2", food2),
    netLogoOutputs += ("final-ticks-food3", food3),
    // The seed is used to control the initialisation of the random number generator of NetLogo
    inputs += seed,
    // Define default values for inputs of the model
    seed := 42,
    gPopulation := 125.0,
    gDiffusionRate := 50.0,
    gEvaporationRate := 50
  )

// Define the hooks to collect the results
val displayHook = ToStringHook(food1, food2, food3)

// Start a workflow with 1 task
val ex = (ants hook displayHook) start
\end{lstlisting}

The code snippet in Listing \ref{lis:openmole-workflow}
introduces several notions from OpenMOLE. First, the original model is
wrapped in a \textit{NetLogoTask}. It is of course not the case for all the different
simulation frameworks. The two other main types of tasks are
the \textit{ScalaTask}, that executes inline Scala code, and the \textit{SystemExecTask},
which runs any kind of application as it would be from a command
line.

The second notion to observe is the Hook \textit{displayHook} associated with the main task. Tasks are mute
pieces of software. They are not conceived to write
files, display values, nor more generally present any
side effects at all. The role of tasks is to compute
some output data from their input data. That's what
guaranties that their execution can be delegated to
other machines.

{\color{black}
OpenMOLE introduces a mechanism called \textit{Hooks} to save or display
results generated on remote environments. Hooks are conceived to
perform an action upon completion of the task they are attached to. In
this example, we use a \textit{ToStringHook} that displays the value of
the task{\textquotesingle}s outputs.}

\subsection{\textbf{Managing the stochasticity}}
{\color{black}
Generally agents models, such as the one we{\textquotesingle}re
studying, are stochastic. It means that their execution depends on the
realisation of random variates. This makes their output variables
random variates as well. These random variates can be studied by
estimating their distribution.}

{\color{black}
Getting one single realisation of the output random variates doesn’t
provide enough information to estimate their distribution. As a
consequence, the model must be executed several times, with different
random sources. All these executions should be statistically
independent to ensure the independent realisation of the
model{\textquotesingle}s output random variates. This operation is
called “replications”.}

{\color{black}
OpenMOLE provides the necessary mechanisms to easily replicate
executions and aggregate the results using a simple statistical
descriptor. The script in Listing \ref{lis:openmole-stochasticity} executes the ants model five times, and
computes the median of each output. The median is a statistical
descriptor of the outputs of the model (however, the form of the
distribution remains unknown).}

{\color{black}
Replicating a stochastic experiment only five times is generally
unreliable. Five is here an arbitrary choice to reduce the global
execution time of this toy example.~}

\begin{lstlisting}[language=scala,label=lis:openmole-stochasticity, caption={\textbf{Median Computation on the Ant Model in OpenMOLE}}]

val modelCapsule = Capsule(ants)

// Define the output variables
val medNumberFood1 = Val[Double]
val medNumberFood2 = Val[Double]
val medNumberFood3 = Val[Double]

// Compute three medians
val statistic =
  StatisticTask() set (
    statistics += (food1, medNumberFood1, median),
    statistics += (food2, medNumberFood2, median),
    statistics += (food3, medNumberFood3, median)
  )

val statisticCapsule = Capsule(statistic)

val seedFactor = seed in (UniformDistribution[Int]() take 5)
val replicateModel = Replicate(modelCapsule, seedFactor, statisticCapsule)

// Define the hooks to collect the results
val displayOutputs = ToStringHook(food1, food2, food3)
val displayMedians = ToStringHook(medNumberFood1, medNumberFood2, medNumberFood3)

// Execute the workflow
val ex = replicateModel + (modelCapsule hook displayOutputs) + (statisticCapsule hook displayMedians) start
\end{lstlisting}

\subsection{\textbf{The optimisation algorithm}}

\textcolor{black}{Now that we have estimators of the output
distribution, we will try to find the parameter settings minimising
these estimators. Listing \ref{lis:openmole-nsga2} describes how to use the NSGA2
multi-objective optimisation algorithm
}\textcolor{black}{\cite{Deb.etal.2002} }\textcolor{black}{in OpenMOLE. The
result files are written
to~}\textrm{\textit{\textcolor{black}{/tmp/ants}}}\textcolor{black}{.~}

\begin{lstlisting}[language=scala,label=lis:openmole-nsga2, caption={\textbf{Parameter Optimisation Using the NSGA-II Genetic Algorithm in OpenMOLE}}]

// Define the population (10) and the number of generations (100).
// Define the inputs and their respective variation bounds.
// Define the objectives to minimize.
// Assign 1 percent of the computing time to reevaluating
// parameter settings to eliminate over-evaluated individuals.
val evolution =
  NSGA2(
    mu = 10,
    termination = 100,
    inputs = Seq(gDiffusionRate -> (0.0, 99.0), gEvaporationRate -> (0.0, 99.0)),
    objectives = Seq(medNumberFood1, medNumberFood2, medNumberFood3),
    reevaluate = 0.01
  )

// Define a builder to use NSGA2 generational EA algorithm.
// replicateModel is the fitness function to optimise.
// lambda is the size of the offspring (and the parallelism level).
val nsga2 =
  GenerationalGA(evolution)(
    replicateModel,
    lambda = 10
  )

// Define a hook to save the Pareto frontier
val savePopulationHook = SavePopulationHook(nsga2, "/tmp/ants/")

// Define another hook to display the generation in the console
val display = DisplayHook("Generation ${" + nsga2.generation.name + "}")

// Plug everything together to create the workflow
val ex = nsga2.puzzle + (nsga2.output hook savePopulationHook hook display) start

\end{lstlisting}

\subsection{\textbf{Scale up}}

When the necessity comes to scale up and expand
the exploration, OpenMOLE's environments come
very handy to quickly distribute the workload of the
workflow to a large computing environment such as the
European Grid Infrastructure (EGI). The optimisation
as we've done so far is not perfectly suited for this
kind of remote environments. In this case, we'll use
the Island model.

Islands are better suited to exploit distributed computing
 resources than classical generational genetic algorithms. 
 Islands of population evolve for a while on
a remote node. When an island is finished, its final
population is merged back into a global archive. A new
island is then generated until the termination criterion
is met: i.e. the total number of islands to generate has been
reached.

{\color{black}
Listing \ref{lis:openmole-islands} shows that implementing islands in the workflow leaves the
script almost unchanged, save for the island and environment
definition. Here we compute 2,000 islands in parallel, each running for 1 hour on the European grid:}

\begin{lstlisting}[language=scala,label=lis:openmole-islands, caption={\textbf{Distribution of the  Parameter Optimisation Process Using the Islands Model}}]

// Define the population (200) and the computation time (1h)
 // The remaining is the same as above
 val evolution =
   NSGA2(
     mu = 200,
     termination = Timed(1 hour),
     inputs = Seq(gDiffusionRate -> (0.0, 99.0), gEvaporationRate -> (0.0, 99.0)),
     objectives = Seq(medNumberFood1, medNumberFood2, medNumberFood3),
     reevaluate = 0.01
   )

 // Define the island model with 2,000 concurrent islands. Each island gets 50 individuals sampled from the global
 // population. The algorithm stops after 200,000 islands evaluations.
 val (ga, island) = IslandSteadyGA(evolution, replicateModel)(2000, 200000, 50)

 val savePopulationHook = SavePopulationHook(ga, "/tmp/ants/")
 val display = DisplayHook("Generation ${" + ga.generation.name + "}")

 // Define the execution environment
 val env = EGIEnvironment("biomed", openMOLEMemory = 1200, wallTime = 4 hours)

 // Define the execution
 val ex =
   (ga.puzzle +
    (island on env) +
(ga.output hook savePopulationHook hook display)) start

\end{lstlisting}

\section{\textbf{Conclusion}}
In this paper, we have shown the features and capabilities of the
OpenMOLE scientific workflow engine.

The light was put on two main components of OpenMOLE: its Domain
Specific Language and the set of distributed environments it can
address. The DSL is an elegant and simple way to describe scientific
workflows from any field of study. The described workflows can then be
executed on a wide range of distributed computing environments
including the most popular job schedulers and grid middlewares.

The DSL and computing environment were then applied to a real-life Ant
simulation model. We showed how to describe a multi-objective
optimisation problem in OpenMOLE, in order to optimise a particular
parameter from the model. The resulting workload was delegated to the
European Grid Infrastructure (EGI).

OpenMOLE as well as all the tools forming its ecosystem are free and
open source software. This allows anyone to contribute to the main
project, or build extensions on top of it.

Future releases of the OpenMOLE platform will integrate a fully
functional web user interface to design workflows, with the DSL still
playing a key part in the design.

\section*{\textbf{Acknowledgment}}
The research leading to these results has received funding from the European Research Council under the European Union’s Seventh Framework Programme (FP/2007-2013) / ERC Grant Agreement n. 319456.

\bibliographystyle{apalike}
\bibliography{HPCS2015-tuto}

\end{document}